\documentclass[universe,article,accept,moreauthors,pdftex]{Definitions/mdpi} 

\firstpage{1} 
\pubvolume{7}
\issuenum{12}
\articlenumber{469}
\pubyear{2021}
\copyrightyear{2021}
\externaleditor{Academic Editors: Stefano Bellucci and Sergei D. Odintsov } %MDPI:please add content of this part
\datereceived{15 November 2021} 
\dateaccepted{27 November 2021} 
\datepublished{30 November 2021} 
\hreflink{https://doi.org/ \linebreak 10.3390/universe7120469} % If needed use \linebreak
\usepackage{latexsym}
\usepackage{graphics}
\usepackage{amssymb}
\usepackage{bm}

\newcommand{\be}[1]{\begin{equation}\label{#1}}
\newcommand{\ee}{\end{equation}}
\newcommand{\ba}[1]{\begin{eqnarray}\label{#1}}
\newcommand{\ea}{\end{eqnarray}}

\newcommand{\nn}{\nonumber}

\Title{Effect of the Cubic Torus Topology on Cosmological~Perturbations}

\TitleCitation{Effect of the Cubic Torus Topology on Cosmological Perturbations}

\Author{Maxim Eingorn %MDPI:Please carefully check the accuracy of names and affiliations.
 $^{1,}$*, Ezgi Canay $^{2}$\orcidA{}, Jacob M. Metcalf $^{1}$,  Maksym Brilenkov $^{3}$ and Alexander Zhuk $^{4}$\orcidB{}}

\AuthorNames{Maxim Eingorn, Ezgi Canay, Jacob M. Metcalf,  Maksym Brilenkov and Alexander Zhuk}

\AuthorCitation{Eingorn, M.; Canay, E.; Metcalf, J.M.; Brilenkov, M.; Zhuk, A.}
\address{%
$^{1}$ \quad Department of Mathematics and Physics, North Carolina Central University, 1801 Fayetteville St., \mbox{Durham, NC 27707, USA}; jmetcal1@eagles.nccu.edu\\
$^{2}$ \quad  Department of Physics, Istanbul Technical University, Maslak, Istanbul 34469, Turkey; ezgicanay@itu.edu.tr\\
$^{3}$ \quad Institute of Theoretical Astrophysics, University of Oslo, Blindern, P.O. Box 1029, N-0315 Oslo, Norway; maksym.brilenkov@astro.uio.no\\
$^{4}$ \quad  Astronomical Observatory, Odessa I.I. Mechnikov National University, Dvoryanskaya St. 2, \mbox{65082 Odessa, Ukraine}; ai.zhuk2@gmail.com}

\corres{Correspondence: maxim.eingorn@gmail.com}

\abstract{We study the effect of the cubic torus topology of the Universe on scalar cosmological perturbations which define the gravitational potential. We obtain three alternative forms of the solution for both the gravitational potential produced by point-like masses, and the corresponding force. The first solution includes the expansion of delta-functions into Fourier series, exploiting periodic boundary conditions. The second one is composed of summed solutions of the Helmholtz equation for the original mass and its images. Each of these summed solutions is the Yukawa potential. In the third formula, we express the Yukawa potentials via Ewald sums. We show that for the present Universe, both the bare summation of Yukawa potentials and the Yukawa-Ewald sums require smaller numbers of terms to yield the numerical values of the potential and the force up to desired accuracy. Nevertheless, the Yukawa formula is yet preferable owing to its much simpler~structure.}


\begin{document}
%%%%%%%%%%%%%%%%%%%%%%%%%%%%%%%%%%%%%%%%%%

\section{Introduction}
\label{sec:1}
Spatial topology of the Universe is the fundamental problem of modern cosmology. Is the Universe spatially flat, open, or closed? Moreover, is it simply or multiply connected? This issue was and is the subject of debate in many scientific articles. Theory, e.g., General Relativity, does not provide direct answers to these questions, hence it is observation instead that plays the decisive role here. 
%Obviously, observations play a central role in solving this %problem. 
For instance, 
%multiple images of a single cosmic object can be seen in %multiply connected space, since photons can travel many times %across its volume~\cite{43,44}. Therefore, 
detection of multiple images of the same object would directly indicate that the space is  multiply connected. Furthermore, such an extended object as the last scattering surface can self-intersect along pairs of circles, the so-called circles-in-the-sky~\cite{45a,45b,45c,45d}. These pairs of matched circles have the same temperature fluctuation distribution. While nearly antipodal circles-in-the-sky have not yet been revealed in the CMB radiation maps, the analysis of CMB anisotropies for repeated patterns is very promising, and even detection of a single pair of matched circles could confirm the flatness of the Universe together with its multi-connectedness~\cite{45,46}. 

At large angular scales, there are observable CMB anomalies in the form of quadrupole moment suppression and the quadrupole and octopole alignment. From the topological point of view, it is natural to explain the suppression by the absence of long wavelengths in sufficiently compact spaces. Such spaces should also have all dimensions of the same order of magnitude, thus being well-proportioned~\cite{37,49}. A cubic torus $T^3$  represents a glaring example of well-proportioned compact space, and a more general rectangular type $T\times T\times  T$  (with unequal periods of tori) can bring forth a symmetry plane or a symmetry axis in the CMB pattern~\cite{50}.

A multiply connected flat space with toroidal topology $T^3$ is compact in all directions and has a finite volume. There are observational limits on the size of such Universe, including the ones coming from the analysis of 7-year and 9-year WMAP temperature maps~\cite{49,52}. For example, according to the 7-year WMAP data, the lower bound on the size of the fundamental topological domain is 27.9 Gpc~\cite{48}. In the case of $T^3$ topology, more recent Planck mission data give  $R_i> 0.92\chi_{\mathrm{rec}}$ and  $R_i> 0.97\chi_{\mathrm{rec}}$ for Planck 2013 and 2015 results, respectively, where $R_i$ is the radius of the largest sphere which can be inscribed in the topological domain,  and $\chi_{\mathrm{rec}}\sim 14\,$Gpc is the distance to the recombination surface~\cite{36,42}. Based on these results, the Planck Collaboration has reported that currently, there is  no detection of compact topology with a characteristic scale being less than the last scattering surface diameter. Meanwhile, as pointed out in~\cite{53}, it is quite possible that the Universe does have compact topology, detectable through the values of observable parameters which lie outside the ranges covered by the WMAP and Planck missions (at least with respect to the circles-in-the-sky search).

Symmetries associated with the cubic torus $T^3$  are inherent in cosmological simulations of the large scale structure formation. Indeed, the cosmological N-body problem is almost always numerically solved in a cubic box with periodic boundary conditions~\cite{19,20,75,76,77,78,79,80,81,82,83}. In view of computational limitations, the edge of the simulation box is smaller than the lower experimental bound on the torus period and ranges from hundreds to thousands of Mpc. To perform such simulations, we need to know the form of metric perturbations, in particular, the expression for the gravitational potential generated by discrete masses. Such an investigation was performed, e.g., in~\cite{71,72} where the Authors considered a toroidal lattice with period $L$ and equal masses $M$ placed at the center of each cell. They found a solution of the Einstein equations, expanded into series in powers of the small parameter $\left(M/L\right)^{1/2}$. It turns out that in this case, the discrete mass distribution is characterized by non-convergent series. The inherent ultraviolet (UV) divergence is related to the point-like nature of the investigated matter sources. In order to avoid 
%the presence of diverging terms in the metric, 
this problem, the Authors provided the masses with a small finite extension, thus introducing a UV cutoff scale. If the Schwarzschild radius of the masses is chosen as this cutoff scale, then one needs the first $10^9$  summands in the series. Meanwhile, if a typical galaxy dimension is chosen instead, then only first 200 summands are needed for an accurate description of the exterior solution.
This cosmological model 
%investigated in~\cite{71,72} 
is characterized by a number of apparent limitations. First of all, clustering in the real Universe is much more complicated and irregular. In the second place, 
%the derived perturbative solution covers only a limited part %of spacetime.  
it is not difficult to show that the obtained solution with point-like gravitating masses has no definite values on
the straight lines joining identical masses in neighboring cells, i.e., at points where masses are absent. The only
way to avoid this problem and get a regular solution
at any point of the cell is to perform the smearing of these masses
over some region, i.e., to employ again a UV cutoff. Exactly the same situation takes place for the gravitational potential as a solution of the Poisson equation with periodic boundary conditions~\cite{BEZ}. 

\textls[-5]{The situation is drastically changed if the gravitational potential satisfies the Helmholtz}-type equation, as it takes place within the cosmic screening approach~\cite{Eingorn1,Claus1,MaximRus,Claus2,MaxEzgi}. Careful analysis of the perturbed Einstein equations reveals that first-order cosmological perturbations (e.g., the gravitational potential) satisfy the Helmholtz equation. This relativistic effect  arises due to the interaction between the gravitational field and the nonzero cosmological background. In the present paper, we analyze this equation in the case of periodic boundary conditions usually assumed for cosmological N-body simulations. In other words, we investigate the impact of the cubic torus topology on the shape of the gravitational potential. 
%We need to know this potential to perform the LSS formation. 
We present three alternative expressions for the potential. 
%  {\cor It is worth noting that these expressions are regular 
%  at all points in space, except for the locations of the 
%  point-like gravitating masses.} 
The main purpose of this paper is to determine among these solutions the one that is most advantageous with respect to numerical applications. Namely, to find which of the solutions requires less terms in series to attain the necessary precision. Our investigation shows that the solution based on Yukawa-type potentials is preferable, provided that the screening length is smaller than the period of the torus. This condition is in agreement with observational bounds. 

The paper is structured as follows. In Section~\ref{sec:2}, we introduce the general setup of the model and present three alternative solutions for the gravitational potential for cubic torus topology. Sections~\ref{sec:3} and \ref{sec:4} are devoted to the detailed study of these potentials and the corresponding forces, respectively, in view of their usefulness for numerical computations. In Section~\ref{sec:5}, we summarize the obtained results. 
 
%%%%%%%%%%%%%%%%%%%%%%%%%%%%%%%%%%%%%%%%%%
\section{The Model and Alternative Solutions}\label{sec:2}
We consider the $\Lambda $CDM model where matter (cold dark and baryonic) is taken in the form of point-like gravitating masses $m_n$. These inhomogeneities perturb the background Friedmann-Lema\^{\i}tre-Robertson-Walker metric. In the conformal Newtonian gauge, the perturbed metric reads~\cite{Mukhanov2,Rubakov}
\be{2.1} ds^2= a^2\left[(1+2\Phi)d\eta^2-(1-2\Phi)\delta_{\alpha\beta}dx^{\alpha}dx^{\beta}\right]\, .
\ee

The first-order scalar perturbation $\Phi$, $|\Phi| \ll 1$, defines the total gravitational potential of the system and satisfies the  %MDPI:Please confirm whether it is necessary to maintain an unindented format
%perturbed Einstein 
equation~\cite{MaxEzgi}
\be{2.2}\Delta\Phi - \frac{a^2}{\lambda^2_{\mathrm{eff}}}\Phi=\frac{\kappa c^2}{2a}\left(\rho -\bar{\rho}\right)\, ,
\ee
where $\kappa\equiv 8\pi G_N/c^4$ ($G_N$ is the Newtonian gravitational constant and $c$ represents the speed of light), $a(\eta)$ is the scale factor, and $\Delta$ denotes the Laplace operator in comoving coordinates. In addition,
\be{2.3}
\rho = \sum_n m_n\delta(\mathbf{r}-\mathbf{r}_n)
\ee
is the comoving mass density and $\bar{\rho}=\rm const$ is its average value. The Helmholtz \mbox{Equation~\eqref{2.2}} was derived in~\cite{MaxEzgi} within the cosmic screening approach~\cite{Eingorn1,Claus1,MaximRus,Claus2} and effectively takes into account peculiar velocities of inhomogeneities. 
The effective screening length
\be{2.4}
\lambda_{\mathrm{eff}}=\sqrt{\frac{c^2a^2H}{3}\int\frac{da}{a^3H^3}}\, ,
\ee
where $H=\left(c/a^2\right)da/d\eta$ is the Hubble parameter. It can be easily seen that $\lambda_{\mathrm{eff}}$ admits the time dependence. If we substitute the cosmological parameters according to the Planck 2018 data~\cite{Planck2018}, i.e., $H_0=67.4 \, {\mathrm {km}}\, {\mathrm {s}}^{-1}{\mathrm {Mpc}}^{-1}$, $\Omega_{\mathrm{M}}=0.315$, $\Omega_{\Lambda}=0.685$, at the present time we get  $(\lambda_{\mathrm{eff}})_0=2.57$ Gpc~\cite{MaxEzgi}.

It is convenient to introduce a shifted gravitational potential

\be{2.5}
\widehat\Phi\equiv \Phi - \lambda^2_{\mathrm{eff}}\frac{\kappa c^2}{2 a^3}\bar\rho\, ,
\ee
which satisfies
\be{2.6}
\Delta\widehat\Phi - \frac{a^2}{\lambda^2_{\mathrm{eff}}}\widehat\Phi=\frac{\kappa c^2}{2a}\rho\, .
\ee

The superposition principle allows us to solve this equation for a selected particle and then, to simply re-express the solution for a system of randomly distributed particles.

Herein we intend to solve this equation in the case of the three-torus  topology $T\times T\times T$ with periods $l_1, l_2$ and $l_3$. Obviously, $\bar\rho = \sum_n m_n/(l_1l_2l_3)$. It is worth noting that the form of the equation is determined by General Relativity with the appropriate choice of the metric and energy-momentum tensor of matter. Therefore, we have the same equation for both flat simply-connected and multiply-connected topologies. Evidently, the solution of this equation depends on the boundary conditions. First, we find the solution for the selected particle $m$, chosen to be, without loss of generality, at the center of Cartesian coordinates.  For the given source, Equation~\eqref{2.6} becomes
\be{2.7}
\Delta\widehat\Phi - \frac{a^2}{\lambda^2_{\mathrm{eff}}}\widehat\Phi=\frac{\kappa c^2}{2a}m\delta (x)\delta (y)\delta (z)\, .
\ee

Toroidal topology also implies periodic boundary conditions.
Therefore, the expansion of the delta-function into Fourier series reads  
\be{2.8} \delta(x)=\frac{1}{l_1}\sum_{k_1=-\infty}^{+\infty}\cos\left(\frac{2\pi k_1}{l_1}x\right)\,  
\ee
and similar expressions apply for $\delta(y)$ and $\delta(z)$. Employing this delta-function presentation, it can be easily verified that the solution of Equation~\eqref{2.7}~is
\ba{2.9} 
\widehat\Phi&=&-\frac{\kappa c^2}{2a}\frac{m}{l_1l_2l_3}\sum_{k_1=-\infty}^{+\infty}\sum_{k_2=-\infty}^{+\infty}
\sum_{k_3=-\infty}^{+\infty}\left[4\pi^2\left(\frac{k_1^2}{l_1^2}+\frac{k_2^2}{l_2^2}+\frac{k_3^2}{l_3^2}\right)+ \frac{a^2}{\lambda^2_{\mathrm{eff}}}\right]^{-1}\,\nn\\
&\times& \cos\left(\frac{2\pi k_1}{l_1}x\right)\cos\left(\frac{2\pi k_2}{l_2}y\right)\cos\left(\frac{2\pi k_3}{l_3}z\right)\, .\,\quad
\ea

Thus, for a system of arbitrarily located massive particles in a cell, the total gravitational potential is
\ba{2.10} 
\widehat\Phi&=&-\frac{\kappa c^2}{2a}\frac{1}{l_1l_2l_3}\sum_n m_n\left\{\sum_{k_1=-\infty}^{+\infty}\sum_{k_2=-\infty}^{+\infty}
\sum_{k_3=-\infty}^{+\infty}\left[4\pi^2\left(\frac{k_1^2}{l_1^2}+\frac{k_2^2}{l_2^2}+\frac{k_3^2}{l_3^2}\right)+ \frac{a^2}{\lambda^2_{\mathrm{eff}}}\right]^{-1} \right.\nn\\
&\times&\left.\cos\left[\frac{2\pi k_1}{l_1}(x-x_n)\right]\cos\left[\frac{2\pi k_2}{l_2}(y-y_n)\right]\cos\left[\frac{2\pi k_3}{l_3}(z-z_n)\right]\right\}\, .\quad
\ea

The obtained solutions satisfy two important natural conditions. First, Equation~\eqref{2.9} yields the correct Newtonian limit in the close vicinity of the source particle. Second, using the relation \eqref{2.5}, it can be demonstrated that the average value of $\Phi$ is equal to zero, as is required of fluctuations at the first-order level. It is worth noting that the sum of Newtonian potentials does not satisfy this condition (see also the reasoning in~\cite{EBV}). 

The solution of Equation~\eqref{2.7} can also be found in the alternative way. Owing to periodic boundary conditions, each mass in the fundamental cell has its counterparts shifted by multiples of tori periods $l_1, l_2$ and $l_3$. Therefore, we may solve Equation~\eqref{2.7} by merely counting the distinct contributions of these images. Since this is a Helmholtz-type equation, the solution is the sum of the corresponding Yukawa potentials:
\ba{2.11} 
\widehat{\Phi}&=&-\frac{\kappa c^2m}{8\pi a}\sum_{k_1=-\infty}^{+\infty}\sum_{k_2=-\infty}^{+\infty}\sum_{k_3=-\infty}^{+\infty} \frac{1}{\sqrt{(x-k_1l_1)^2+(y-k_2l_2)^2+(z-k_3l_3)^2}}\nn\\
&\times&\exp\left(-\frac{a\sqrt{(x-k_1l_1)^2+(y-k_2l_2)^2+(z-k_3l_3)^2}}{\lambda_{\mathrm{eff}}}\right)\, .
\ea

We rewrite the alternative solutions \eqref{2.9} and \eqref{2.11} as
%
% start a new page without indent 4.6cm
%\clearpage
\end{paracol}
\nointerlineskip
\ba{2.12} 
\tilde\Phi_{\cos}\equiv\left(-\frac{G_N m}{c^2al}\right)^{-1}\widehat\Phi_{\cos}&=&\sum_{k_1=-\infty}^{+\infty}\sum_{k_2=-\infty}^{+\infty}
\sum_{k_3=-\infty}^{+\infty}\left[\pi\left({k_1^2}+{k_2^2}+{k_3^2}\right)+ \frac{1}{4\pi\tilde\lambda^2_{\mathrm{eff}}\,}\right]^{-1}\,\nn\\
&\times&\cos\left({2\pi k_1}\tilde x\right)\cos\left({2\pi k_2}\tilde y\right)\cos\left({2\pi k_3}\tilde z\right)\,\nn\\ 
\ea
\begin{paracol}{2}
%\linenumbers
\switchcolumn

%%%%%%%
and
%%%%%%
% start a new page without indent 4.6cm
%\clearpage
\end{paracol}
\nointerlineskip
\ba{2.13} 
\tilde{\Phi}_{\exp}\equiv\left(-\frac{G_N m}{c^2al}\right)^{-1}\widehat{\Phi}_{\exp}
&=&\sum_{k_1=-\infty}^{+\infty}\sum_{k_2=-\infty}^{+\infty}\sum_{k_3=-\infty}^{+\infty} \frac{1}{\sqrt{(\tilde x-k_1)^2+(\tilde y-k_2)^2+(\tilde z-k_3)^2}}\nn\\
&\times&\exp\left(-\frac{\sqrt{(\tilde x-k_1)^2+(\tilde y-k_2)^2+(\tilde z-k_3)^2}}{\tilde\lambda_{\mathrm{eff}}}\right)\, , 
\ea
\begin{paracol}{2}
%\linenumbers
\switchcolumn
\noindent where, for simplicity, we consider an equal-sided cubic torus with $l_1=l_2=l_3\equiv l$ and introduce the notation
\be{2.14} 
x=\tilde xl,\quad y=\tilde yl,\quad z=\tilde zl,\quad \lambda_{\mathrm{eff}}=\tilde\lambda_{\mathrm{eff}}al\, .
\ee

Yukawa potentials with periodic boundaries can also be expressed in 
the form of Ewald sums, i.e., as two distinct rapidly converging 
series, each of which exists in one of the real and Fourier spaces. 
This type of presentation is usually used in depicting electrostatic 
interactions in plasma, colloids etc., and for this purpose, the Yukawa 
potential for systems with three-dimensional periodicity was obtained 
earlier in~\cite{Salin}. In the cosmological framework, the Yukawa-Ewald 
potential for gravitational interactions takes the~form
\ba{2.15}
\tilde{\Phi}_{\mathrm{mix}}&\equiv&\left(-\frac{G_N m}{c^2al}\right)^{-1}\widehat{\Phi}_{\mathrm{mix}}\,\nn\\
&=&\sum_{k_1=-\infty}^{+\infty}\sum_{k_2=-\infty}^{+\infty}\sum_{k_3=-\infty}^{+\infty}\left\{\frac{D\left(\sqrt{(\tilde x-k_1)^2+(\tilde y-k_2)^2+(\tilde z-k_3)^2};\alpha;\tilde\lambda_{\mathrm{eff}}\right)}{2\sqrt{(\tilde x-k_1)^2+(\tilde y-k_2)^2+(\tilde z-k_3)^2}}\right. \,\nn\\
&+&\left.4\pi \cos\left[2\pi \left(k_1\tilde x+k_2\tilde y+k_3\tilde z\right) \right]\frac{\exp\left[-\left(4\pi^2k^2+\tilde\lambda^{-2}_{\mathrm{eff}}\right)/\left(4\alpha^2\right)\right]}{4\pi^2k^2+\tilde\lambda^{-2}_{\mathrm{eff}}}\right\} \, ,
\ea
where $k^2\equiv k_1^2+k_2^2+k_3^2$,
%%%%%%
\ba{2.16}
&&D\left(\sqrt{(\tilde x-k_1)^2+(\tilde y-k_2)^2+(\tilde z-k_3)^2};\alpha;\tilde\lambda_{\mathrm{eff}}\right)\,\nn\\
&\equiv&\exp\left(\frac{\sqrt{(\tilde x-k_1)^2+(\tilde y-k_2)^2+(\tilde z-k_3)^2}}{\tilde\lambda_{\mathrm{eff}}}\right)\,\nn\\
&\times&\mathrm{erfc}\left(\alpha \sqrt{(\tilde x-k_1)^2+(\tilde y-k_2)^2+(\tilde z-k_3)^2}+\frac{1}{2\alpha\tilde\lambda_{\mathrm{eff}}}\right)\,\nn\\
&+&\exp\left(-\frac{\sqrt{(\tilde x-k_1)^2+(\tilde y-k_2)^2+(\tilde z-k_3)^2}}{\tilde\lambda_{\mathrm{eff}}}\right)\,\nn\\
&\times&\mathrm{erfc}\left(\alpha \sqrt{(\tilde x-k_1)^2+(\tilde y-k_2)^2+(\tilde z-k_3)^2}-\frac{1}{2\alpha\tilde\lambda_{\mathrm{eff}}}\right)\, .
\ea

In these expressions, erfc represents the complementary error function and $\alpha$, the 
free parameter, is to be assigned the optimal value to save 
computational effort while operating with adequate accuracy. Below we will test a number of values of $\alpha$.
%  , choosing them as multiples of the screening length
%  $\tilde\lambda_{\mathrm{eff}}$.
Our research demonstrates that for the chosen range of $\tilde\lambda_{\mathrm{eff}}$, the optimal one is around $2$.
%%%%%%%%%%%%%%

We note that alternative expressions for the gravitational potential were also found in the cases of slab and chimney topologies~\cite{slab,chimney}. Direct comparison of the obtained formulas for different topologies shows that only the Formula \eqref{2.13} above (Yukawa potentials) may be interpreted as a simple extension of the ``slab'' and ``chimney'' counterparts (2.28) in~\cite{slab} and (18) in~\cite{chimney}, respectively. However, obviously, both Formulas \eqref{2.12} and \eqref{2.15} are quite different from their counterparts, and this is a nontrivial task to derive them from the previous papers. Moreover, the Yukawa-Ewald potential is not presented in the case of slab topology. Thus, formulas found in the present paper are new and otherwise absent in the literature. As regards the gravitational forces derived below, again, the similarity is present only in the case of the Yukawa formula, but both alternatives are new and cannot be easily derived from the previous results on different topologies.

{{The obtained solutions \eqref{2.12}, \eqref{2.13} and \eqref{2.15} depend on time. It is important to note that they all satisfy the same Helmholtz equation and are three representations of the solution. In our papers~\cite{Eingorn1,Claus1,MaximRus,MaxEzgi} it was investigated in detail that such a solution satisfies the complete system of perturbed Einstein equations.}}
%%%%%%%%%%%%%%%%%%%%%%%%%%%%%%%%%%%%%%%%%%
\section{Gravitational Potentials}
\label{sec:3} 
In the previous section, we have obtained  three alternative formulas for the gravitational potential created by a point-like particle placed at the center of Cartesian coordinates $(x,y,z) =(0,0,0)$ and by its infinitely many images placed at points $(x,y,z)=(k_1l,k_2l,k_3l)$ where $k_{1,2,3}=0,\pm 1,\pm2,\ldots\,$ Due to periodic boundary conditions, these formulas include infinite series.  We aim to find out which of these expressions requires fewer terms in the series sum to yield the value of the potential to given accuracy. The less the number $n$ of these terms, the more advantageous the corresponding formula for numerical calculations. This number $n$ is defined via the condition that the ratio $|\mathrm{exact}\ \tilde\Phi - \mathrm{approximate}\ \tilde\Phi|/|\mathrm{exact}\ \tilde\Phi|$ is either equal to or less than 0.001. This is our demanded level of precision in determining the approximate value of $\tilde\Phi$. Each of the alternative expressions \eqref{2.12}, \eqref{2.13} and \eqref{2.15} has its own number $n$ designated as $n_{\mathrm{cos}}, n_{\mathrm{exp}}$ and $n_{\mathrm{mix}}$, correspondingly. Evidently, the formula with the smallest number is, all other things being equal, the most convenient for numerical computations. All three formulas to be compared contain triple series. Consequently, the sought values $n$ correspond to the minimum number of triplets $(k_1, k_2,k_3)$ included in series for which the required precision is achieved. To find this number, we generate a sequence in increasing order of $\sqrt{k_1^2+k_2^2+k_3^2}$ in Mathematica~\cite{Math} and count the number $n$ of terms involved in it.

We calculate the potentials \eqref{2.12}, \eqref{2.13} and \eqref{2.15} in Mathematica~\cite{Math} up to the adopted accuracy at a selection of points in the cell and display the results in Tables~\ref{results_table_1} and~\ref{results_table_2}. The number $n_{\mathrm{exp}}$ is defined employing Equation~\eqref{2.13}: for any $n>n_{\mathrm{exp}}$, the approximate  
$\tilde{\Phi}_{\exp}$ will be determined with better accuracy than a tenth of a percent. The values $n_{\mathrm{cos}}$ and $n_{\mathrm{mix}}$ follow from the Formulas \eqref{2.12} and \eqref{2.15} under the condition that the gravitational potential is calculated with the same accuracy at the point of interest. We have found that the Yukawa-Ewald formula \eqref{2.15}  works well (i.e., requires the smallest number of terms) both for small and large selected values of the screening length $\tilde\lambda_{\mathrm{eff}}$. Therefore, we evaluate the exact $\tilde{\Phi}$ by the Formula \eqref{2.15} for $n\gg n_{\mathrm{mix}}$. 
Additionally, we have observed that the use of Equation~\eqref{2.12} to get $n_{\mathrm{cos}}$ yields faulty outputs due to problematic aspects of the computational process. The ``trigonometric'' Formula~\eqref{2.12} contains an alternating series. The summation of such a series is accompanied by significant round-off errors and to reach the required accuracy, when possible, it is necessary to take into account a very large number of terms (more than $10^5$ in our case). Therefore, the use of this formula looks absolutely unreasonable in comparison with the rapidly converging expressions \mbox{\eqref{2.13} and \eqref{2.15}}. Hence, this trigonometric formula is not suitable for numerical calculations, and the related values are excluded from the tables. 

As follows from the Formulas \eqref{2.12}, \eqref{2.13} and \eqref{2.15}, the resulting values of the potential are sensitive to the choice of  $\tilde\lambda_{\mathrm{eff}}$. For our calculations, we choose four different values, that are $\tilde\lambda_{\mathrm{eff}} = 0.01, 0.1, 1$ and $5$. The rescaled screening length $\tilde\lambda_{\mathrm{eff}}$ is the ratio  of the physical effective screening length $\lambda_{\mathrm{eff}}$ to the physical size of the period $al$.
As we have mentioned previously, today $\lambda_{\mathrm{eff}}\sim 2.6$ Gpc for  the $\Lambda$CDM model~\cite{MaxEzgi}, and the size of the fundamental domain for the cubic torus topology is restricted by Planck 2015 results to no less than $al \sim$ 27 Gpc~\cite{42}, i.e., the observational data require that $\tilde\lambda_{\mathrm{eff}} \ll 1$. Nevertheless, taking into account that many N-body simulations are indeed performed in boxes with sizes less than 1 Gpc, we also consider $\tilde\lambda_{\mathrm{eff}} \geq 1$.

The Yukawa-Ewald potential \eqref{2.15} is sensitive to the free parameter $\alpha$ as well. Therefore, choosing different values of this quantity, we also seek those at which $n_{\mathrm{mix}}$ will be minimum. Our calculations demonstrate that for the chosen range of $0.01\leq\tilde\lambda_{\mathrm{eff}}\leq 5$, the optimal value of $\alpha$ is around 2. 

\end{paracol}
\begin{specialtable}[H] 
    \tablesize{\small}
    \widetable
\caption{Rescaled potential $\tilde\Phi$ and the corresponding numbers $n_{\exp}$ of terms in the series sum at nine points in the cell for $\tilde\lambda_{\mathrm{eff}}=0.01$ (left chart) and $\tilde\lambda_{\mathrm{eff}}=0.1$ (right chart).} \label{results_table_1}

\setlength{\cellWidtha}{\columnwidth/12-2\tabcolsep-0.1in}
\setlength{\cellWidthb}{\columnwidth/12-2\tabcolsep-0.1in}
\setlength{\cellWidthc}{\columnwidth/12-2\tabcolsep-0.1in}
\setlength{\cellWidthd}{\columnwidth/12-2\tabcolsep-0.1in}
\setlength{\cellWidthe}{\columnwidth/12-2\tabcolsep+0.5in}
\setlength{\cellWidthf}{\columnwidth/12-2\tabcolsep-0.1in}
\setlength{\cellWidthg}{\columnwidth/12-2\tabcolsep-0.1in}
\setlength{\cellWidthh}{\columnwidth/12-2\tabcolsep-0.1in}
\setlength{\cellWidthi}{\columnwidth/12-2\tabcolsep-0.1in}
\setlength{\cellWidthj}{\columnwidth/12-2\tabcolsep-0.1in}
\setlength{\cellWidthk}{\columnwidth/12-2\tabcolsep+0.5in}
\setlength{\cellWidthl}{\columnwidth/12-2\tabcolsep-0.1in}
\scalebox{1}[1]{\begin{tabularx}{\columnwidth}{>{\PreserveBackslash\centering}m{\cellWidtha}>{\PreserveBackslash\centering}m{\cellWidthb}>{\PreserveBackslash\centering}m{\cellWidthc}>{\PreserveBackslash\centering}m{\cellWidthd}>{\PreserveBackslash\centering}m{\cellWidthe}>{\PreserveBackslash\centering}m{\cellWidthf}>{\PreserveBackslash\centering}m{\cellWidthg}>{\PreserveBackslash\centering}m{\cellWidthh}>{\PreserveBackslash\centering}m{\cellWidthi}>{\PreserveBackslash\centering}m{\cellWidthj}>{\PreserveBackslash\centering}m{\cellWidthk}>{\PreserveBackslash\centering}m{\cellWidthl}}
	\toprule
\multicolumn{6}{c}{\boldmath$\tilde\lambda_{\mathrm{eff}}=0.01$} & \multicolumn{6}{c}{\boldmath$\tilde\lambda_{\mathrm{eff}}=0.1$} \\\midrule
	& \boldmath$\tilde x$ & \boldmath$\tilde y$ &\boldmath$\tilde z$& \boldmath$\tilde{\Phi}$ & \boldmath$n_{\exp}$&	& \boldmath$\tilde x$ & \boldmath$\tilde y$ &\boldmath$\tilde z$& \boldmath$\tilde{\Phi}$ & \boldmath$n_{\exp}$ \\\midrule

	$A_1$ & 0.5 &  0 & 0.5 & $1.105\times10^{-30}$ & 9&	$A_1$ & 0.5 & 0 & 0.5 & $4.837\times10^{-3}$ & 20\\

	$A_2$ & 0.5 & 0 & 0.1 & $2.810\times10^{-22}$ & 2&	$A_2$ & 0.5 & 0 & 0.1 & $2.406\times10^{-2}$ & 9  \\

	$A_3$ & 0.5 &  0 & 0 & $7.715\times10^{-22}$ & 2 &$A_3$ & 0.5 & 0 & 0 & $2.705\times10^{-2}$ & 9 \\

	$B_1$ & 0.1 & 0 & 0.1 & $5.101\times10^{-6}$ & 1 &	$B_1$ & 0.1 & 0 & 0.1 & $1.719$ & 1\\

	$B_2$ & 0.1 & 0 & 0 & $4.540\times10^{-4}$ & 1&	$B_2$ & 0.1 & 0 & 0 & $3.679$ & 1\\

	$C_1$ & 0.5 & 0.5 & 0.5 & $2.262\times10^{-37}$ & 20& $C_1$ & 0.5 & 0.5 & 0.5 & $1.602\times 10^{-3}$ & 20 \\

	$C_2$ & 0.5 & 0.5 & 0.1 & $5.413\times10^{-31}$ & 8& $C_2$ & 0.5 & 0.5 & 0.1 & $4.478\times10^{-3}$ & 20\\

	$C_3$ & 0.1 & 0.5 & 0.1 & $1.044\times10^{-22}$ & 3&$C_3$ & 0.1 & 0.5 & 0.1 & $2.146\times 10^{-2}$ & 12\\
	
	$C_4$ & 0.1 & 0.1 & 0.1 & $1.735\times10^{-7}$ & 1&$C_4$ & 0.1 & 0.1 & 0.1 & $1.022$ & 1 \\
	
	\bottomrule
\end{tabularx}}
\end{specialtable}
\begin{paracol}{2}
%\linenumbers
\switchcolumn
\vspace{-9pt}

%\end{paracol}

\begin{specialtable}[H] 
    \tablesize{\small}
    %\widetable
\caption{Rescaled potential $\tilde\Phi$ and the corresponding numbers $n_{\exp}$ and $n_{\mathrm{mix}}$ of terms in the series sum at nine points for  $\tilde\lambda_{\mathrm{eff}}=1$ (top chart) and $\tilde\lambda_{\mathrm{eff}}=5$ (bottom chart).} \label{results_table_2}

\setlength{\cellWidtha}{\columnwidth/9-2\tabcolsep+0.0in}
\setlength{\cellWidthb}{\columnwidth/9-2\tabcolsep+0.0in}
\setlength{\cellWidthc}{\columnwidth/9-2\tabcolsep+0.0in}
\setlength{\cellWidthd}{\columnwidth/9-2\tabcolsep+0.0in}
\setlength{\cellWidthe}{\columnwidth/9-2\tabcolsep+0.0in}
\setlength{\cellWidthf}{\columnwidth/9-2\tabcolsep+0.0in}
\setlength{\cellWidthg}{\columnwidth/9-2\tabcolsep+0.0in}
\setlength{\cellWidthh}{\columnwidth/9-2\tabcolsep+0.0in}
\setlength{\cellWidthi}{\columnwidth/9-2\tabcolsep+0.0in}
\scalebox{1}[1]{\begin{tabularx}{\columnwidth}{>{\PreserveBackslash\centering}m{\cellWidtha}>{\PreserveBackslash\centering}m{\cellWidthb}>{\PreserveBackslash\centering}m{\cellWidthc}>{\PreserveBackslash\centering}m{\cellWidthd}>{\PreserveBackslash\centering}m{\cellWidthe}>{\PreserveBackslash\centering}m{\cellWidthf}>{\PreserveBackslash\centering}m{\cellWidthg}>{\PreserveBackslash\centering}m{\cellWidthh}>{\PreserveBackslash\centering}m{\cellWidthi}}
	\toprule
	& \boldmath$\tilde x$ & \boldmath$\tilde y$ & \boldmath$\tilde z$ &\boldmath$\tilde{\Phi}$ & \boldmath$n_{\exp}$ & \boldmath$n^{\alpha=1}_{\mathrm{mix}}$ & \boldmath$n^{\alpha=2}_{\mathrm{mix}}$& \boldmath$n^{\alpha=3}_{\mathrm{mix}}$\\\midrule

	$A_1$ & 0.5 & 0 & 0.5 & $12.00$ & 3449 & 64 & 9& 25  \\

		$A_2$ & 0.5 & 0 & 0.1 & $12.42$ & 3352 & 48 & 7&22\\

		$A_3$ & 0.5 & 0 & 0 & $12.47$ & 3345 &47& 7&22\\

		$B_1$ & 0.1 & 0 & 0.1 & $16.77$ & 2965 & 37 & 7&23 \\

		$B_2$ & 0.1 & 0 & 0 & $19.66$ & 2794 & 33 & 7&24   \\

		$C_1$ & 0.5 & 0.5 & 0.5 & $11.79$ & 3510 & 64 & 20& 24  \\

		$C_2$ & 0.5 & 0.5 & 0.1 & $11.98$ & 3451 & 62 & 8& 24 \\

		$C_3$ & 0.1 & 0.5 & 0.1 & $12.37$ & 3370 & 54 & 7& 21\\

		$C_4$ & 0.1 & 0.1 & 0.1 & $15.50$ & 3058 & 39 &7& 22 \\\midrule

	& \boldmath$\tilde x$ & \boldmath$\tilde y$ & \boldmath$\tilde z$ &\boldmath$\tilde{\Phi}$ & \boldmath$n^{\alpha=1}_{\mathrm{mix}}$ & \boldmath$n^{\alpha=2}_{\mathrm{mix}}$ & \boldmath$n^{\alpha=2.5}_{\mathrm{mix}}$& \boldmath$n^{\alpha=5}_{\mathrm{mix}}$\\\midrule

		$A_1$ & 0.5 & 0 & 0.5 & $313.6$ & 14 & 1 & 1& 17  \\

		$A_2$ & 0.5 & 0 & 0.1 & $314.0$ & 11 & 2 & 1&16\\

		$A_3$ & 0.5 & 0 & 0 & $314.1$ & 11 &2& 1&17\\

		$B_1$ & 0.1 & 0 & 0.1 & $318.4$ & 12 & 1 & 3&23 \\

		$B_2$ & 0.1 & 0 & 0 & $321.3$ & 12 & 1 & 4&29    \\

		$C_1$ & 0.5 & 0.5 & 0.5 & $313.4$ & 20 & 1 & 1& 12  \\

		$C_2$ & 0.5 & 0.5 & 0.1 & $313.6$ & 13 & 1 & 1& 6 \\

		$C_3$ & 0.1 & 0.5 & 0.1 & $314.0$ & 13 & 2 & 1& 11\\

		$C_4$ & 0.1 & 0.1 & 0.1 & $317.2$ & 12 & 1 &3& 17 \\
	
	\bottomrule
\end{tabularx}}
\end{specialtable}
%
%\begin{paracol}{2}
%%\linenumbers
%\switchcolumn
%
In Table~\ref{results_table_1}, we give the minimum numbers $n_{\exp}$ of terms in the series \eqref{2.13} that return the value of the gravitational potential up to the adopted accuracy. The left and right charts correspond to $\tilde\lambda_{\mathrm{eff}}=0.01$ and $\tilde\lambda_{\mathrm{eff}}=0.1$, respectively. As for the values $n_{\mathrm{mix}}$, these numbers are the same as $n_{\exp}$
for both left and right charts as long as $\alpha$ lies between $10^{-3}$ and $2$. However, outside this interval, the Yukawa-Ewald formula may require more summands. For example, if $\tilde\lambda_{\mathrm{eff}}=0.1$ and $\alpha=2.5$, then $n_{\mathrm{mix}}=29,23,23,4,1,48,28,23,6$ for the points $A_1,A_2,\ldots ,C_4$, respectively. All in all, when $\tilde\lambda_{\mathrm{eff}} \ll 1$ (in accordance with observational bounds), both the Yukawa \eqref{2.13} and Yukawa-Ewald \eqref{2.15} formulas demonstrate good results since they need less terms in the series sum. The  potential expression in \eqref{2.13} is much simpler, though. Thus, from that aspect, the Yukawa formula is a more practical tool for computational purposes in the case of small $\tilde\lambda_{\mathrm{eff}}$.  

In Table~\ref{results_table_2}, we present the results of similar calculations for $\tilde\lambda_{\mathrm{eff}}=1$ and $\tilde\lambda_{\mathrm{eff}}=5$. In the case $\tilde\lambda_{\mathrm{eff}}=1$ (top chart), $n_{\exp}\gg n_{\mathrm{mix}}$ for all selected points, and the optimal choice for the parameter $\alpha$ in the Yukawa-Ewald formula \eqref{2.15} is 2. For $\tilde\lambda_{\mathrm{eff}}=5$ (bottom chart), as regards the Yukawa formula \eqref{2.13}, the $n_{\exp}$ values are ill-suited (they are extremely large). Therefore, we do not show them here. The value $\alpha =2$ is again the optimal choice for the Yukawa-Ewald potential. Hence, when $\tilde\lambda_{\mathrm{eff}} \geq 1$, the Yukawa-Ewald formula \eqref{2.15} delivers the best performance in numerical calculations.

We demonstrate in Figures \ref{fig:1} and \ref{fig:2} (plotted in Mathematica~\cite{Math}) the $\tilde z=0$ sections of the rescaled potential $\tilde\Phi$ for different values of the rescaled effective screening length $\tilde\lambda_{\mathrm{eff}}$ considered in Tables~\ref{results_table_1} and ~\ref{results_table_2}. To plot both figures, we use the Yukawa-Ewald formula \eqref{2.15} for $\alpha=2$ and $n \gg n_{\mathrm{mix}}$.

\newpage
\end{paracol}
\begin{figure}[H]	
\widefigure
\includegraphics[scale=0.35]{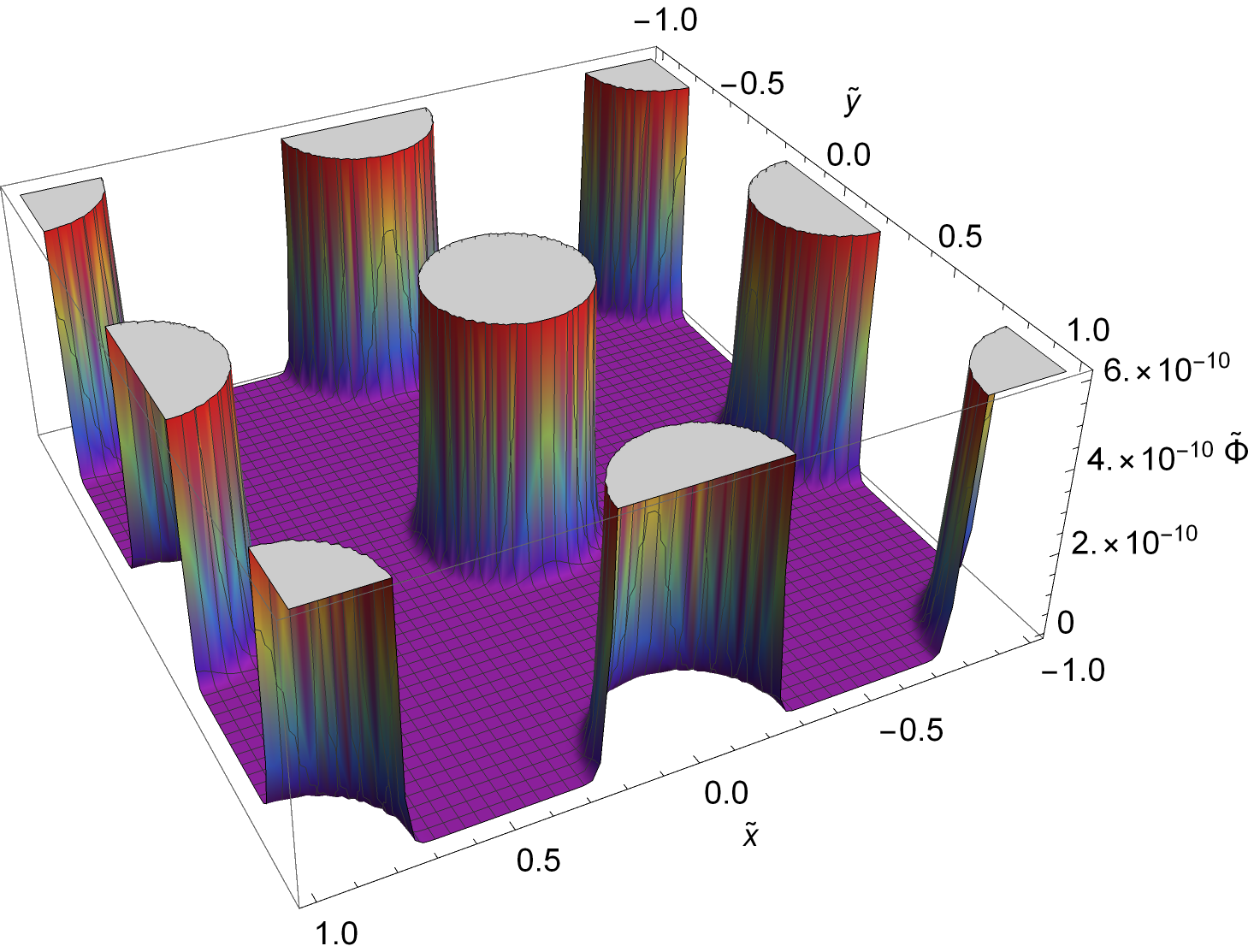}\hspace*{0.1in}
\includegraphics[scale=0.35]{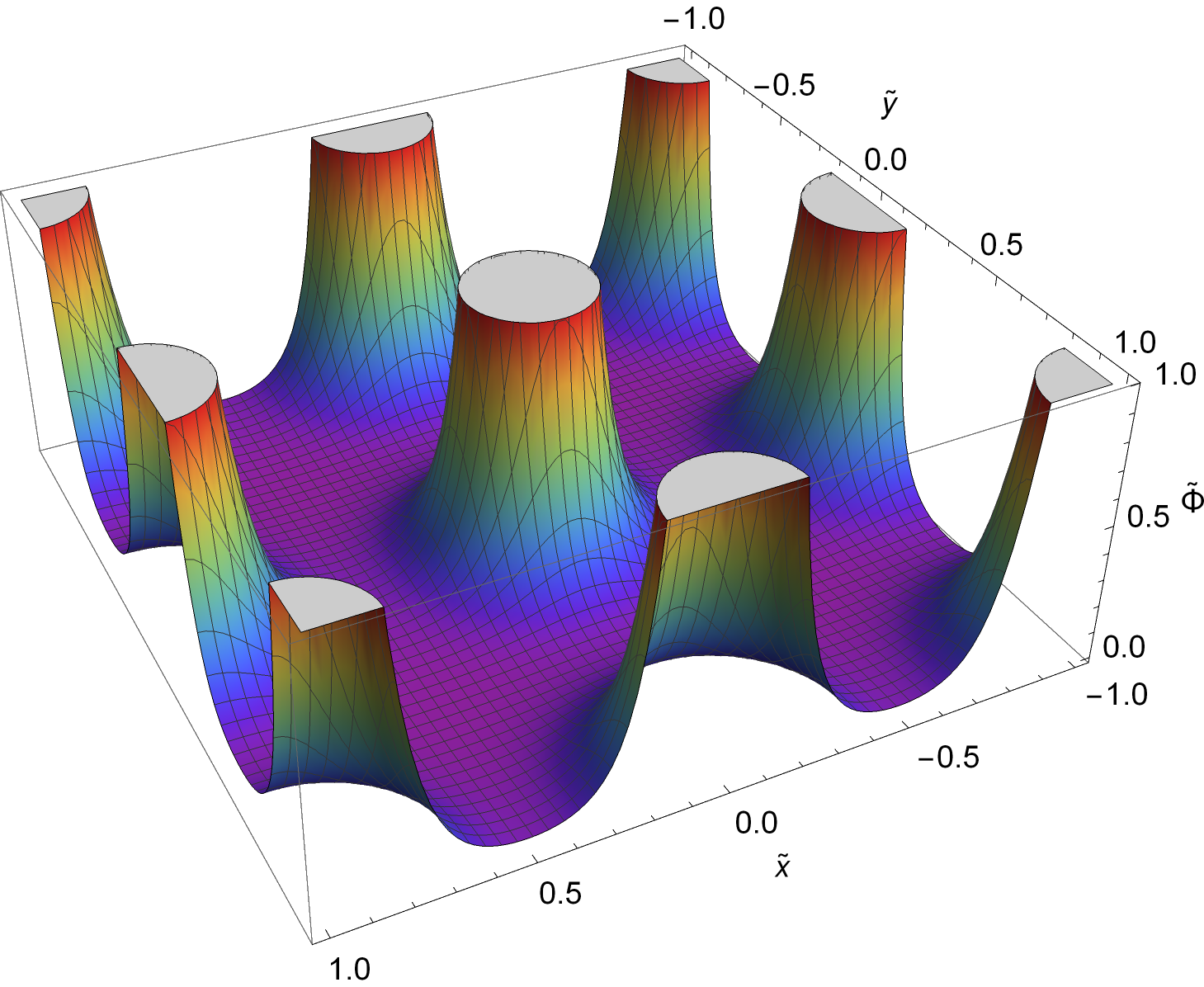}
%\resizebox{0.47\textwidth}{!}{\includegraphics{}}\quad\quad
%	\resizebox{0.47\textwidth}{!}{\includegraphics{}}
\caption{$\tilde z=0$ sections of the rescaled  potential  $\tilde\Phi=\left[-G_N m/(c^2al)\right]^{-1}\widehat\Phi$ for $\tilde\lambda_{\mathrm{eff}} = 0.01$ (\textbf{left panel}) and $\tilde\lambda_{\mathrm{eff}} = 0.1$ (\textbf{right panel}), respectively.\label{fig:1}}
\end{figure}  
\vspace{-6pt}

\begin{figure}[H]	
\widefigure
\includegraphics[scale=0.35]{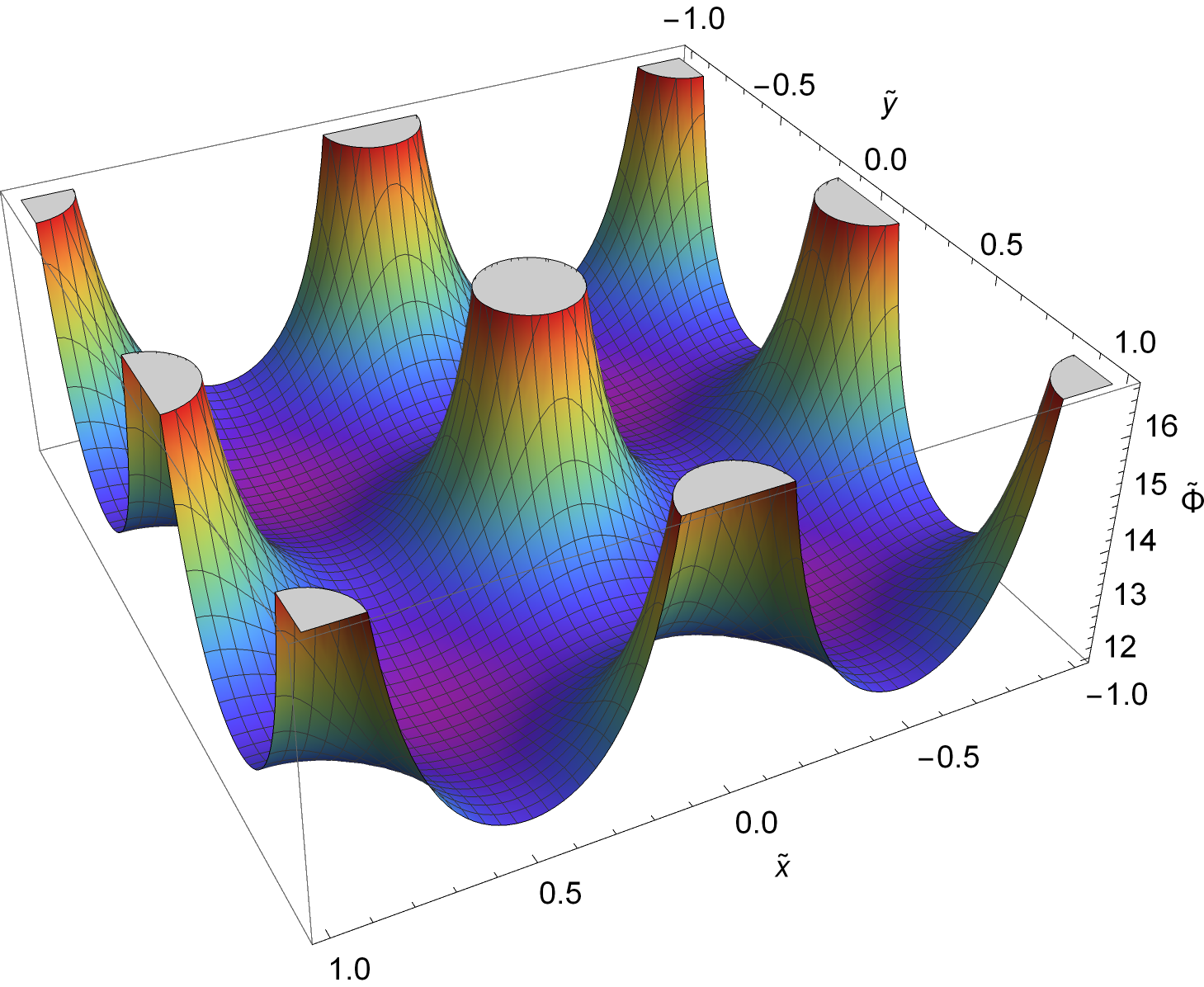}\hspace*{0.1in}
\includegraphics[scale=0.35]{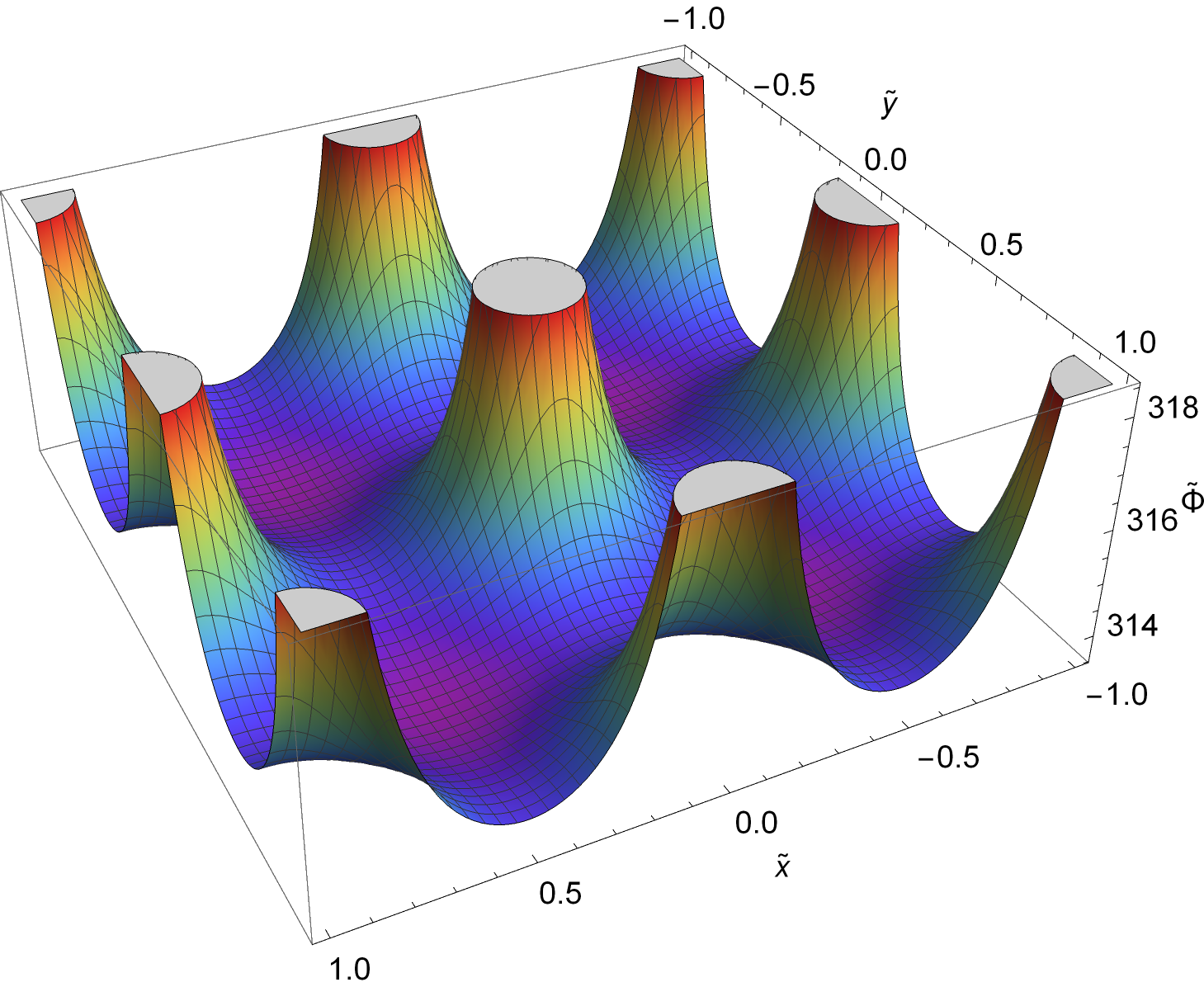}
%\resizebox{0.47\textwidth}{!}{\includegraphics{}}\quad\quad
%	\resizebox{0.47\textwidth}{!}{\includegraphics{}}
\caption{$\tilde z=0$ sections of the rescaled  potential  $\tilde\Phi=\left[-G_N m/(c^2al)\right]^{-1}\widehat\Phi$ for $\tilde\lambda_{\mathrm{eff}} = 1$ (\textbf{left panel}) and $\tilde\lambda_{\mathrm{eff}} = 5$ (\textbf{right panel}), respectively.\label{fig:2}}
\end{figure}
\begin{figure}[H]	
\widefigure
\includegraphics[scale=0.35]{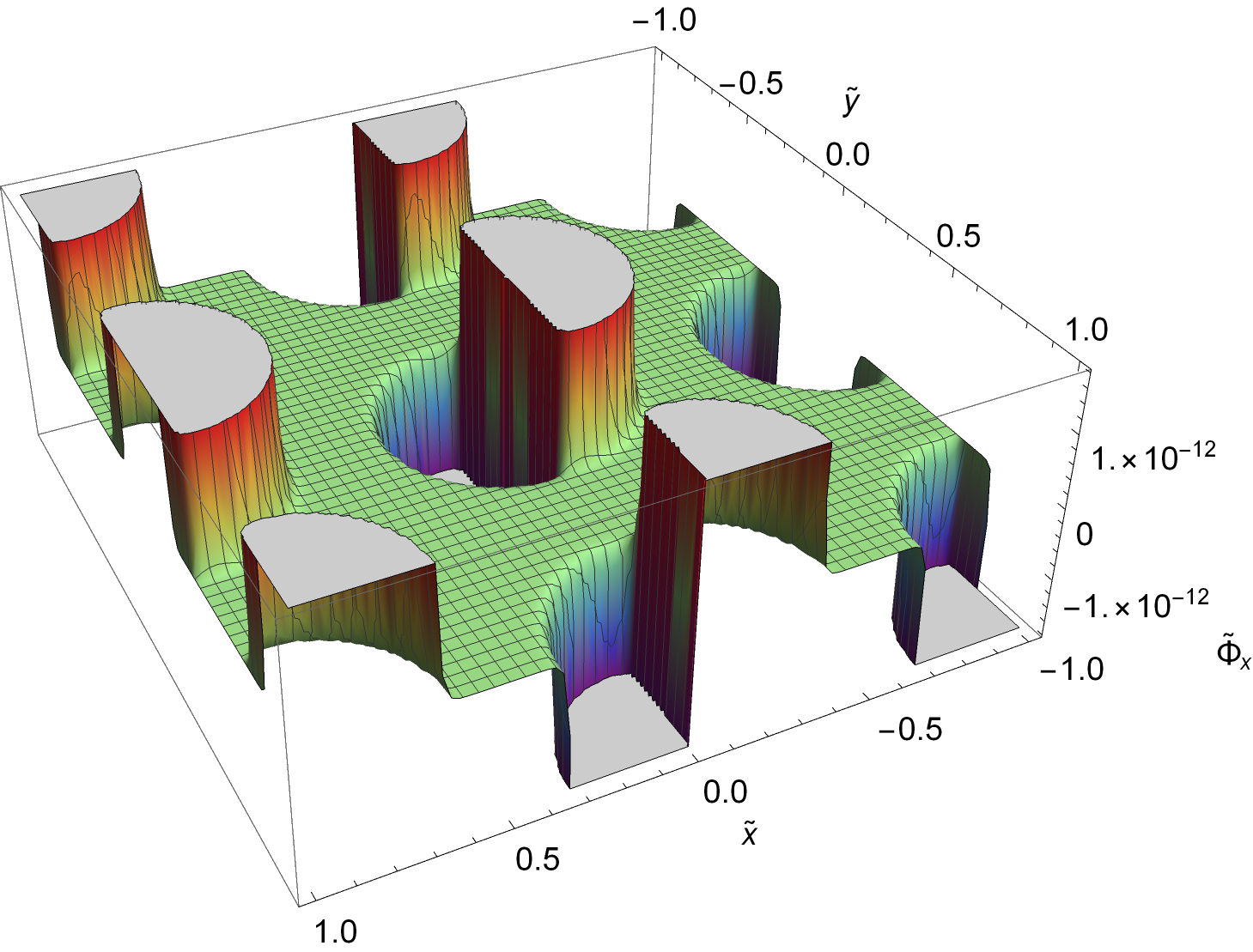}\hspace*{0.1in}
\includegraphics[scale=0.35]{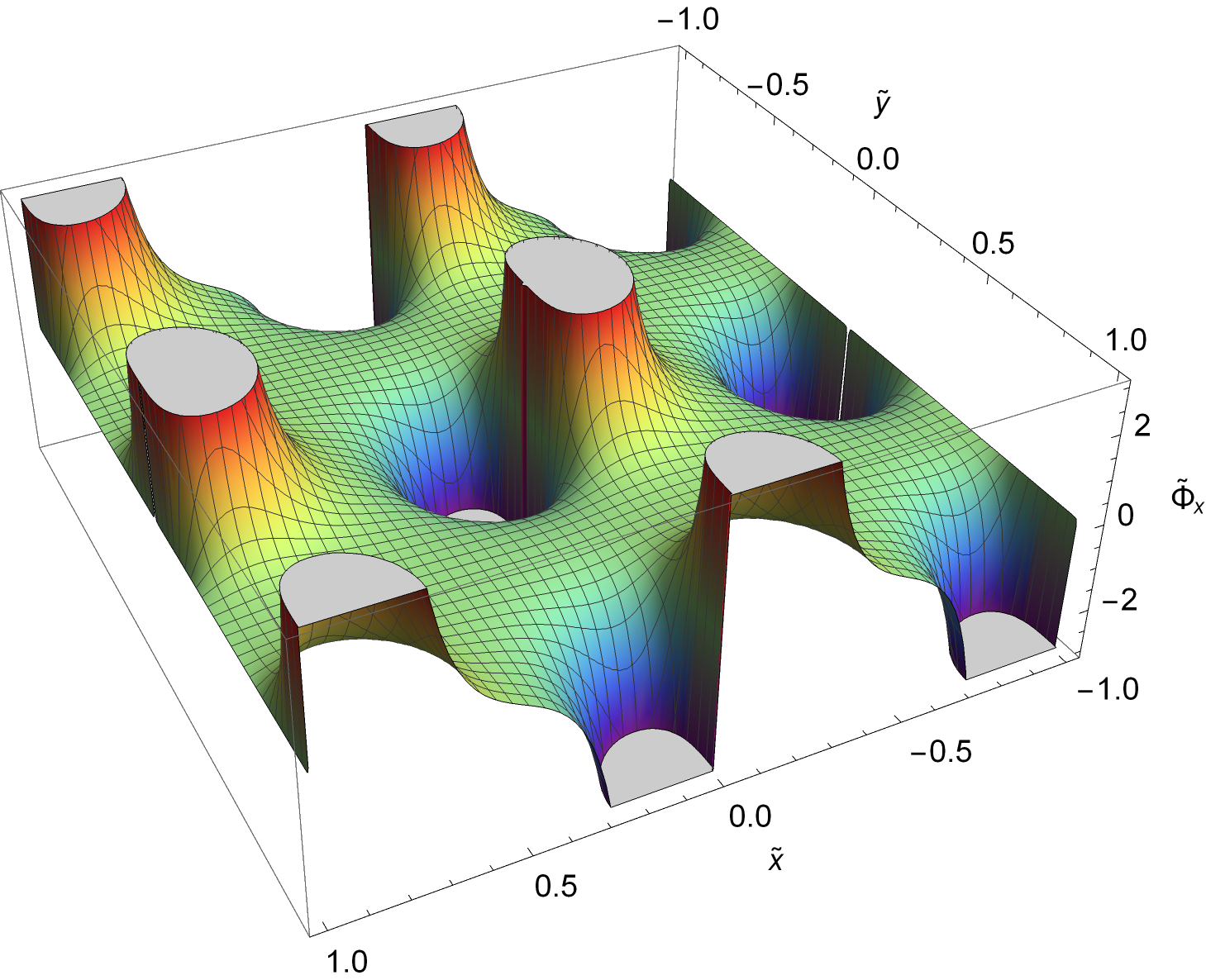}

%\resizebox{0.48\textwidth}{!}{\includegraphics{}}\quad\quad
%	\resizebox{0.48\textwidth}{!}{\includegraphics{}}
\caption{$x$-component of the force,  $\tilde{\Phi}_x\equiv\partial \tilde{\Phi}/\partial\tilde x$, for $\tilde\lambda_{\mathrm{eff}} = 0.01$  (\textbf{left panel}) and $\tilde\lambda_{\mathrm{eff}} = 0.1$ (\textbf{right panel}).\label{fig:3}}
\end{figure}

  
\begin{paracol}{2}
%\linenumbers
\switchcolumn

\section{Gravitational Forces}\label{sec:4}

In this section we provide the gravitational force formulas associated with the alternative expressions \eqref{2.12}, \eqref{2.13} and \eqref{2.15} for the gravitational potential. Clearly, it is sufficient to consider the force projection onto one of the axes in Cartesian coordinates. Let it be the $x$-axis. Then, the projections read
%%%%%
\ba{4.1} 
\frac{\partial}{\partial\tilde x}\left(\tilde{\Phi}_{\cos}\right)&=&\ -2\pi \sum_{k_1=-\infty}^{+\infty}\sum_{k_2=-\infty}^{+\infty}
\sum_{k_3=-\infty}^{+\infty}\left[\pi\left({k_1^2}+{k_2^2}+{k_3^2}\right)+ \frac{1}{4\pi\tilde\lambda^2_{\mathrm{eff}}\,}\right]^{-1}\,\nn\\
&\times&\ k_1\sin\left({2\pi k_1}\tilde x\right)\cos\left({2\pi k_2}\tilde y\right)\cos\left({2\pi k_3}\tilde z\right)\, ,\,\quad
\ea
\newpage
%%%%%%%
%%%%%%%
% start a new page without indent 4.6cm
%\clearpage
\end{paracol}
\nointerlineskip
\ba{4.2} 
&&\frac{\partial}{\partial\tilde x}\left(\tilde{\Phi}_{\exp}\right)=-\sum_{k_1=-\infty}^{+\infty}\sum_{k_2=-\infty}^{+\infty}\sum_{k_3=-\infty}^{+\infty} \exp\left(-\frac{\sqrt{(\tilde x-k_1)^2+(\tilde y-k_2)^2+(\tilde z-k_3)^2}}{\tilde\lambda_{\mathrm{eff}}}\right)\nn\\
&\times&\left\{\frac{\tilde x-k_1}{\left[(\tilde x-k_1)^2+(\tilde y-k_2)^2+(\tilde z-k_3)^2\right]^{3/2}}+\vphantom{\frac{\tilde x-k_1}{\left[(\tilde x-k_1)^2+(\tilde y-k_2)^2+(\tilde z-k_3)^2\right]^{3/2}}}\frac{\tilde x-k_1}{\tilde\lambda_{\mathrm{eff}}\left[(\tilde x-k_1)^2+(\tilde y-k_2)^2+(\tilde z-k_3)^2\right]}\right\}\, ,\quad\ea
\begin{paracol}{2}
%\linenumbers
\switchcolumn
\vspace{-10pt}

%%%%%%%
% start a new page without indent 4.6cm
%\clearpage
\end{paracol}
\nointerlineskip
\ba{4.3}
&&\frac{\partial}{\partial\tilde x}\left(\tilde{\Phi}_{\mathrm{mix}}\right)=\,\nn\\&&-\frac{1}{2}\sum_{k_1=-\infty}^{+\infty}\sum_{k_2=-\infty}^{+\infty}\sum_{k_3=-\infty}^{+\infty}\left\{\frac{\left(\tilde x-k_1\right) D\left(\sqrt{(\tilde x-k_1)^2+(\tilde y-k_2)^2+(\tilde z-k_3)^2};\alpha;\tilde\lambda_{\mathrm{eff}}\right)}{[(\tilde x-k_1)^2+(\tilde y-k_2)^2+(\tilde z-k_3)^2]^{3/2}}\right.\nn\\
&+&C_-\frac{\left(\tilde x-k_1\right)\exp\left(-\sqrt{(\tilde x-k_1)^2+(\tilde y-k_2)^2+(\tilde z-k_3)^2}/\tilde\lambda_{\mathrm{eff}}\right)}{(\tilde x-k_1)^2+(\tilde y-k_2)^2+(\tilde z-k_3)^2}\,\nn\\
&+&C_+\frac{\left(\tilde x-k_1\right)\exp\left(\sqrt{(\tilde x-k_1)^2+(\tilde y-k_2)^2+(\tilde z-k_3)^2}/\tilde\lambda_{\mathrm{eff}}\right)}{(\tilde x-k_1)^2+(\tilde y-k_2)^2+(\tilde z-k_3)^2}\,\nn\\
&+&\left.
16\pi^2  k_1\sin\left[2\pi\left(k_1\tilde x+k_2\tilde y+k_3\tilde z\right)\right]\frac{\exp\left[-\left(4\pi^2k^2+\tilde\lambda^{-2}_{\mathrm{eff}}\right)/\left(4\alpha^2\right)\right]}{4\pi^2k^2+\tilde\lambda^{-2}_{\mathrm{eff}}}\right\}\, ,
\ea
\begin{paracol}{2}
%\linenumbers
\switchcolumn

%%%%%%
\noindent where
%%%%%%
\ba{4.4}
C_{\mp}&=&C_{\mp}\left(\sqrt{(\tilde x-k_1)^2+(\tilde y-k_2)^2+(\tilde z-k_3)^2};\alpha;\tilde\lambda_{\mathrm{eff}}\right)\,\nn\\
&\equiv&\frac{2\alpha}{\sqrt{\pi}}\exp\left[-\left(\alpha\sqrt{(\tilde x-k_1)^2+(\tilde y-k_2)^2+(\tilde z-k_3)^2}\mp\frac{1}{2\alpha\tilde\lambda_{\mathrm{eff}}}\right)^2\right]\,\nn\\
&\pm&\frac{1}{\tilde\lambda_{\mathrm{eff}}}\mathrm{erfc}\left(\alpha\sqrt{(\tilde x-k_1)^2+(\tilde y-k_2)^2+(\tilde z-k_3)^2}\mp\frac{1}{2\alpha\tilde\lambda_{\mathrm{eff}}}\right)\, .
\ea
%%%%%%%
%  for the trigonometric \eqref{2.12}, Yukawa \eqref{2.13} 
%  and Yukawa-Ewald \eqref{2.15} formulas, respectively. 

The $x$-components of the force are zero at the points $A_1,A_2,A_3,C_1$ and $C_2$, so we calculate these components up to the adopted accuracy only at the points $B_1,B_2,C_3$ and $C_4$.

The results we arrive at are as follows: first, as also is the case for potentials, the trigonometric Formula \eqref{4.1} does not provide acceptable values because of the complications that arise during the computational stage.
Next, if $\tilde\lambda_{\mathrm{eff}}=0.01$, then $n_{\exp}=1,1,3,1$ for the points $B_1,B_2,C_3,C_4$, respectively, and the numbers $n_{\mathrm{mix}}$ are exactly the same (assuming here and in what follows that $\alpha=2$). In the case $\tilde\lambda_{\mathrm{eff}}=0.1$, we get $n_{\exp}=1,1,12,1$ for these points, and the corresponding numbers $n_{\mathrm{mix}}$ are, again, identical. However, for $\tilde\lambda_{\mathrm{eff}}=1$, we have $n_{\exp}=258,82,987,486$ while $n_{\mathrm{mix}}=7,7,21,9$. Finally, if $\tilde\lambda_{\mathrm{eff}}=5$, then still $n_{\mathrm{mix}}=7,7,21,9$, but $n_{\exp}$ acquires unreasonably large values. 

Our calculations demonstrate that the numbers $n_{\exp}$ start to grow once $\tilde\lambda_{\mathrm{eff}}$ exceeds 0.1 and they acquire large values as $\tilde\lambda_{\mathrm{eff}}$ approaches 1, while $n_{\mathrm{mix}}$ remain small throughout (provided that the value of the parameter $\alpha$ is optimal, e.g., for $\alpha=2$ in the above cases). The Yukawa formula is, after all, an attractive option in view of its simpler structure. The Yukawa-Ewald formula is much more complicated and consequently, it takes longer to numerically calculate the potentials and forces for comparable values of  $n_{\mathrm{exp}}$ and  $n_{\mathrm{mix}}$. There exists a moment, though, when the execution time for the Yukawa formula, with increased number of terms in the sum, is approximately equal to the one for  the more complex Yukawa-Ewald formula. We have seen that with respect to gravitational forces, this takes place when $n_{\mathrm{exp}}$ is about 6 times larger than $n_{\mathrm{mix}}$.

In Figures~\ref{fig:3} and \ref{fig:4} (plotted in Mathematica~\cite{Math}), we depict the $\tilde z=0$ sections of the $x$-components of gravitational forces for  $\tilde\lambda_{\mathrm{eff}}=0.01$, $0.1$ and $1$.
We employ the Yukawa-Ewald formula \eqref{4.3} for $n \gg n_{\mathrm{mix}}$ and $\alpha=2$. For $ \tilde\lambda_{\mathrm{eff}}=5$, the picture is similar to one in the case $\tilde\lambda_{\mathrm{eff}}=1$.
\end{paracol}

\begin{paracol}{2}
%\linenumbers
\switchcolumn
\vspace{-6pt}

\begin{figure}[H]	
\includegraphics[scale=0.4]{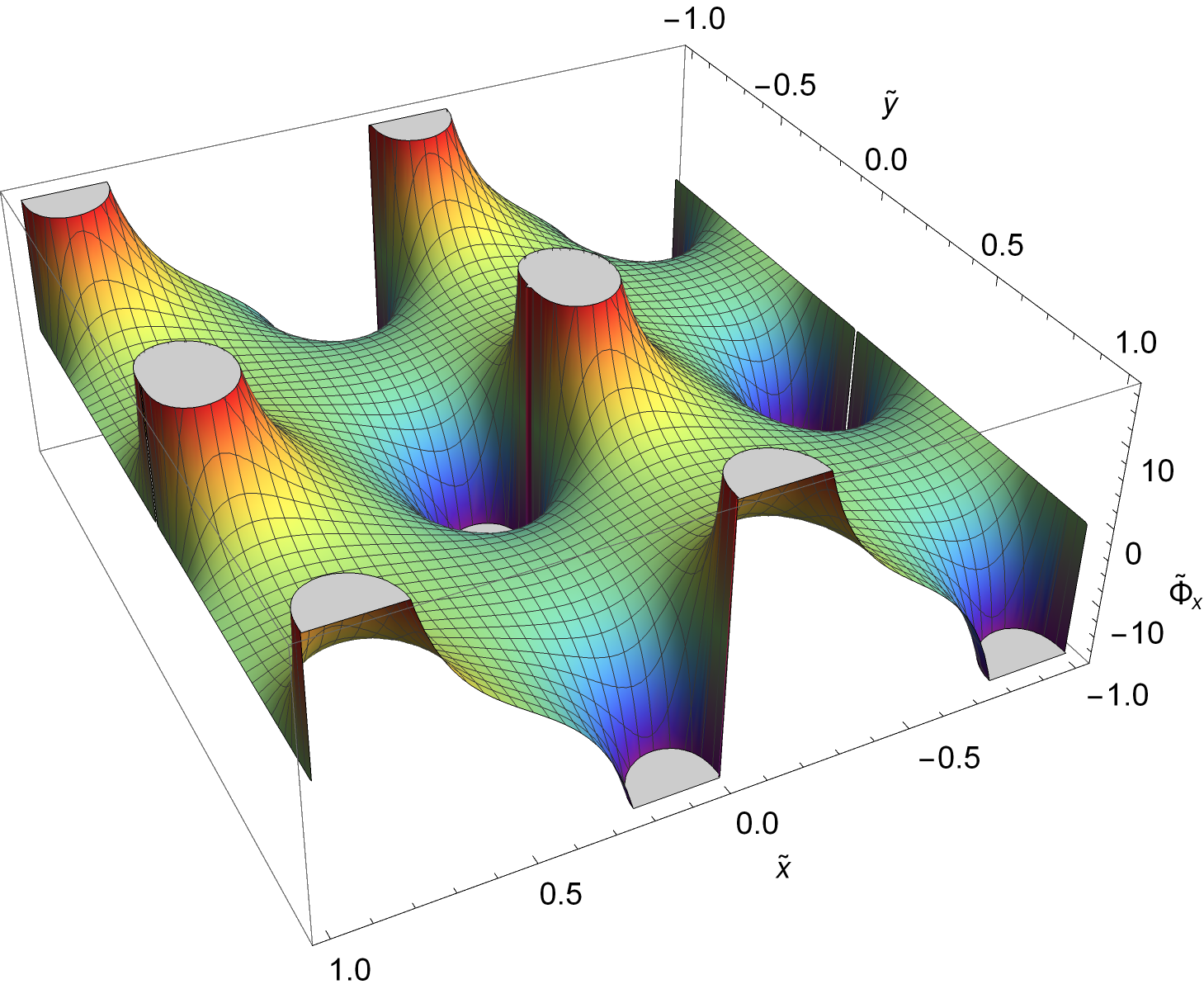}
%\resizebox{0.58\textwidth}{!}{\includegraphics{force3TTT}}
\caption{$x$-component of the force,  $\tilde{\Phi}_x\equiv\partial \tilde{\Phi}/\partial\tilde x$, for $\tilde\lambda_{\mathrm{eff}} = 1$.\label{fig:4}}
\end{figure}  
\newpage
%%%%%%%%%%%%%%%%%%%%%%%%%%%%%%%%%%%%%%%%%%
\section{Conclusions}\label{sec:5}
In this paper we have analyzed the influence of topology on the gravitational interaction in the Universe. We have considered a model in which the fundamental domain is a three-torus $T\times T\times T$. In such a space, gravitating masses are subject to periodic boundary conditions along three coordinate axes, that is, every mass in the fundamental domain has its counterparts in infinitely many cells shifted along each axis by multiples of tori periods. For this lattice Universe, we have obtained three alternative forms of the expression for the gravitational potential produced by a point-like mass. The first one (see Equation~\eqref{2.12}) exploits the periodic structure of space: it involves the expansion of delta-functions into Fourier series. This solution is a trigonometric series, thus we named it trigonometric. 
The second one (see Equation~\eqref{2.13}), the Yukawa solution, was obtained by directly summing the fields produced by the original mass and its images. Since the summed potentials satisfy the Helmholtz equation, they are the Yukawa potentials. In the third formula (see Equation~\eqref{2.15}), we have expressed them via Ewald sums (the Yukawa-Ewald formula) and shown that in some cases, such a trick facilitates (despite the complex form of the resulting expression) numerical calculations. We have also presented the corresponding formulas for the $x$-component of the gravitational force (see Equations~\eqref{4.1}--\eqref{4.3}). All these formulas, both for the potentials and forces, can be easily generalized for a system of arbitrarily located massive particles (see, e.g., \eqref{2.10}). It is well known that the gravitational potential of a system of masses distributed in the Universe defines the scalar perturbations of the~metric.

A reasonable question to ask is, then, which of these formulas is indeed preferable for numerical applications for the given accuracy. Since all three expressions are sensitive to the rescaled effective  screening length $\tilde\lambda_{\mathrm{eff}}=\lambda_{\mathrm{eff}}/(al)$, we analyzed both small values and values equal to or greater than 1: $0.01\leq \tilde\lambda_{\mathrm{eff}}\leq 5$. The Yukawa-Ewald formula additionally admits a free parameter $\alpha$, and we have revealed that for the given range of $\tilde\lambda_{\mathrm{eff}}$, the optimal value of $\alpha$ is 2.
Our calculations show that the trigonometric formula does not provide reasonable results for the potentials or forces as complications arise in the computational process. In the case of small $\tilde\lambda_{\mathrm{eff}}$ (as also demanded by the observational bounds) both the Yukawa and Yukawa-Ewald formulas deliver good results since they require rather small numbers of terms $n_{\mathrm{exp}}$ and  $n_{\mathrm{mix}}$ to yield the values for the potentials and forces up to the given accuracy. Nevertheless, employing the Yukawa formula is a more convenient choice owing to its notably simpler structure. The situation is altered when $\tilde\lambda_{\mathrm{eff}} > 0.1$: $n_{\mathrm{exp}}$  begins to increase quickly while $n_{\mathrm{mix}}$ still takes on rather small values. Therefore, for such $\tilde\lambda_{\mathrm{eff}}$, the Yukawa-Ewald formula is preferable instead.

Finally, we emphasize two important points. First, our results directly confirm that the undesirable impact of periodicity on simulation outputs can be weakened if the edge of the box (cubic torus period $al$) is set to be larger than the predicted Yukawa interaction range $\lambda_{\mathrm{eff}}$ (see Table~\ref{results_table_1}). The Yukawa formula reflects the contribution of images to the value of the potential via the number $n_{\mathrm{exp}}$ (needed to reach the required accuracy). We can easily see that  $n_{\mathrm{exp}}$ gets smaller (i.e., goes to 1) with decreasing $\tilde\lambda_{\mathrm{eff}}$. Second, operating with summed Yukawa potentials, we  provide a reliable description of the inhomogeneous gravitational field generated by a toroidal lattice of point-like masses, avoiding non-convergent series. The obtained series converge at all points only except those where discrete masses themselves are located.
\vspace{6pt}

%%%%%%%%%%%%%%%%%%%%%%%%%%%%%%%%%%%%%%%%%%
\authorcontributions{Conceptualization, M.E.; methodology, M.E., E.C. and A.Z.; formal analysis, M.E., E.C., J.M.M., M.B. and A.Z.; investigation, M.E., E.C., J.M.M., M.B. and A.Z.; writing---original draft preparation, A.Z.; writing---review and editing, M.E. and E.C.; visualization, M.E. and J.M.M.; supervision, M.E. and A.Z.; project administration, M.E.; funding acquisition, M.E. All authors have read and agreed to the published version of the manuscript.}

\funding{The work of Maxim Eingorn and Jacob M. Metcalf was supported by the National Science Foundation HRD Award number 1954454.}

\conflictsofinterest{The authors have no conflict of interest to declare that are relevant to the content of this article.} 

%%%%%%%%%%%%%%%%%%%%%%%%%%%%%%%%%%%%%%%%%%
\end{paracol}
\reftitle{References}

% If authors have biography, please use the format below
%\section*{Short Biography of Authors}
%\bio
%{\raisebox{-0.35cm}{\includegraphics[width=3.5cm,height=5.3cm,clip,keepaspectratio]{Definitions/author1.pdf}}}
%{\textbf{Firstname Lastname} Biography of first author}
%
%\bio
%{\raisebox{-0.35cm}{\includegraphics[width=3.5cm,height=5.3cm,clip,keepaspectratio]{Definitions/author2.jpg}}}
%{\textbf{Firstname Lastname} Biography of second author}

% The following MDPI journals use author-date citation: Arts, Econometrics, Economies, Genealogy, Humanities, IJFS, JRFM, Laws, Religions, Risks, Social Sciences. For those journals, please follow the formatting guidelines on http://www.mdpi.com/authors/references
% To cite two works by the same author: \citeauthor{ref-journal-1a} (\citeyear{ref-journal-1a}, \citeyear{ref-journal-1b}). This produces: Whittaker (1967, 1975)
% To cite two works by the same author with specific pages: \citeauthor{ref-journal-3a} (\citeyear{ref-journal-3a}, p. 328; \citeyear{ref-journal-3b}, p.475). This produces: Wong (1999, p. 328; 2000, p. 475)

\end{document}